\documentstyle[12pt]{article}
\def\singlespace {\smallskipamount=3.75pt plus1pt minus1pt
                  \medskipamount=7.5pt plus2pt minus2pt
                  \bigskipamount=15pt plus4pt minus4pt
                  \normalbaselineskip=15pt plus0pt minus0pt
                  \normallineskip=1pt
                  \normallineskiplimit=0pt
                  \jot=3.75pt
                  {\def\smallskip {\vskip\smallskipamount}}
                  {\def\medskip   {\vskip\medskipamount}}
                  {\def\bigskip   {\vskip\bigskipamount}}
                  {\setbox\strutbox=\hbox{\vrule
                    height10.5pt depth4.5pt width 0pt}}
                  \parskip 7.5pt
                  \normalbaselines}
\def\middlespace {\smallskipamount=5.625pt plus1.5pt minus1.5pt
                  \medskipamount=11.25pt plus3pt minus3pt
                  \bigskipamount=22.5pt plus6pt minus6pt
                  \normalbaselineskip=22.5pt plus0pt minus0pt
                  \normallineskip=1pt
                  \normallineskiplimit=0pt
                  \jot=5.625pt
                  {\def\smallskip {\vskip\smallskipamount}}
                  {\def\medskip   {\vskip\medskipamount}}
                  {\def\bigskip   {\vskip\bigskipamount}}
                  {\setbox\strutbox=\hbox{\vrule
                    height15.75pt depth6.75pt width 0pt}}
                  \parskip 11.25pt
                  \normalbaselines}
\def\doublespace {\smallskipamount=7.5pt plus2pt minus2pt
                  \medskipamount=15pt plus4pt minus4pt
                  \bigskipamount=30pt plus8pt minus8pt
                  \normalbaselineskip=30pt plus0pt minus0pt
                  \normallineskip=2pt
                  \normallineskiplimit=0pt
                  \jot=7.5pt
                  {\def\smallskip {\vskip\smallskipamount}}
                  {\def\medskip   {\vskip\medskipamount}}
                  {\def\bigskip   {\vskip\bigskipamount}}
                  {\setbox\strutbox=\hbox{\vrule
                    height21.0pt depth9.0pt width 0pt}}
                  \parskip 15.0pt
                  \normalbaselines}

\begin{document}

\def\lsim{\:\raisebox{-0.5ex}{$\stackrel{\textstyle<}{\sim}$}\:}
\def\gsim{\:\raisebox{-0.5ex}{$\stackrel{\textstyle>}{\sim}$}\:}

\begin{flushright}
PRL-TH-94/08 \\
TIFR/TH/94-07 \\
\end{flushright}
\begin{center}
\large {\bf CONSTRAINTS ON BARYON-NONCONSERVING YUKAWA COUPLINGS IN A
SUPERSYMMETRIC THEORY} \\
\bigskip
Biswajoy Brahmachari$^{\rm (a)}$\footnote {Address after 1st
October 1994: High Energy Section, International Center For
Theoretical Physics, P.O. Box 586, 34100 Trieste, Italy.} and
Probir Roy$^{\rm (b)}$ \\
\end{center}

\begin{enumerate}

\item[{(a)}] Theory Group, Physical Research Laboratory, Navrangpura, \\
Ahmedabad 380 009, INDIA, biswajoy@prl.ernet.in \\

\item[{(b)}] Theoretical Physics Group, Tata Institute of Fundamental
Research,  \\ Homi Bhabha Road, Bombay 400 005, INDIA,
probir@tifrvax.bitnet \\

\end{enumerate}

\noindent PACS number(s) : 11.30.Pb,~12.10.Dm

{\singlespace
\begin{center}
\underbar{Abstract} \\
\end{center}

The 1-loop evolution of couplings in the minimal supersymmetric
standard model, extended to include baryon nonconserving $(B\!\!\!/)$
operators through explicit $R$-parity violation, is considered keeping only
$B\!\!\!/$ superpotential terms involving the maximum possible number
of third generation superfields. If all retained Yukawa couplings $Y_i$
are required to remain in the perturbative domain $(Y_i < 1)$ upto the
scale of gauge group unification, upper bounds ensue on
the magnitudes of the $B\!\!\!/$ coupling strengths at the
supersymmetry breaking scale, independent of the model of unification.
They turn out to be similar to the corresponding fixed point
values reached from a wide range of $Y_i$ (including all $Y_i$ greater
than unity) at the unification scale.   The coupled evolution of the
top and $B\!\!\!/$ Yukawa couplings results in a reduction of the
fixed point value of the former.

}

\newpage
\middlespace
There is widspread interest today in baryon number violating
$(B\!\!\!/)$ processes.  Many studies have been made of $B\!\!\!/$
couplings in the context of baryogenesis, proton decay etc.  Yet, one
of the least investigated sources of baryon nonconservation is the set
of $B\!\!\!/$ operators which occur in the superpotential of the
minimal supersymmetric standard model (MSSM) extended to include
explicit $R$-parity breaking terms [1].

In this paper we start with all such operators coming from the
superpotential. We know from the Standard Model that Yukawa coupling
strengths spectacularly reflect the generation hierarchy.  The third
generation ones are the strongest -- followed by those of the second
generation with the first generation couplings being the weakest.  It
would be reasonable to adopt a similar hypothesis for the $B\!\!\!/$
Yukawa couplings and retain only those which are the strongest --
discarding the rest.  Because of color antisymmetry, not all three
superfields in the trilinear superpotential term can belong to the
third generation.  We deem it sufficient to retain only
those terms in which the third generation superfields appear twice.
We bound the
magnitudes of the corresponding coupling strengths from above by
utilizing constraints from Renormalization Group Evolution (RGE).  The
bounds result from the requirement that all Yukawa couplings in the
theory remain in the perturbative domain upto the scale of the
unification of gauge couplings [2] $M_U \sim 2 \times 10^{16}$ GeV.

$R$-parity $R_p \equiv (-)^{3B+L+2S}$ (with $B,L,S$ as
baryon number, lepton number and spin respectively) distinguishes
between particles $(R_p = 1)$ and sparticles $(R_p = -1)$.  The popular
formulation [3] of the MSSM has $R_p$-conservation built into it by
fiat.  Many recent studies suggest [4] nonetheless $R\!\!\!/_p$
couplings, allowed by renormalizability and supersymmetry, that admit both
$B$ and $L$ violation.  However, the simultaneous presence of both
types of terms and the (naturalness-based) supposition that the
supersymmetry breaking scale $M_{SUSY}$ should be $\lsim 0$ (TeV) would
imply catastrophic proton decay unobserved in nature.  This forces
practitioners of explicit $R\!\!\!/_p$ models to consider {\it either}
lepton {\it or} baryon nonconserving cases.

Thus two $R\!\!\!/_p$ scenarios are in vogue.  There is one with purely lepton
number violating interactions, respecting baryon number:

\noindent \underbar{Scenario 1} : ${\cal L}_{L\!\!\!/} = \lambda_{ijk}
(L_iL_j \bar E_k)_F + \lambda'_{ijk} (L_i Q_j \bar D_k)_F$. \hfill (1)
\\
The second $R\!\!\!/_p$ secnario has purely baryon number
violating terms with conserved lepton number:

\noindent \underbar{Scenario 2} : ${\cal L}_{B\!\!\!/} =
\lambda^{\prime\prime}_{ijk} [\bar D_i \bar D_j \bar U_k]_F$. \hfill
(2) \\
Here $\lambda,\lambda'$ and $\lambda^{\prime\prime}$ are Yukawa
couplings with $\lambda'$ and $\lambda^{\prime\prime}$ being
antisymmetric in $i,j$.  Moreover, $L,Q,\bar E,\bar D,\bar U$ stand for
the doublet lepton, doublet quark, singlet antilepton, singlet
$d$-antiquark, and singlet $u$-antiquark superfields respectively,
with subscripts acting as generation indices.

Either (1) or (2) has been shown [4] to be consistent with some deeper
fundamental theory and therefore can be seriously entertained.
Moreover, cosmological bounds on $\lambda,\lambda'$ and
$\lambda^{\prime\prime}$, once believed to be strong [5], have been
shown [6] to be not generally valid.  In particular, a GUT era
leptogenesis makes $\lambda^{\prime\prime}_{ijk}$ completely free [6]
of cosmological constraints.  Serious phenomenological upper bounds do
exist [7], of course, on the magnitudes of most of the $\lambda$ and
$\lambda'$ components from the nonobservation of various leptonic
rare decays, neutrino Majorana mass etc. Comparatively, the
$\lambda^{\prime\prime}$-components, apart from
$\lambda^{\prime\prime}_{211}$ and $\lambda^{\prime\prime}_{311}$ [8]
which have been strongly bounded from the lack of any observed $n\bar
n$
oscillation, stand relatively unconstrained.  It is therefore
worthwhile to try to derive some upper bounds, however weak, on the
magnitudes of the remaining $\lambda^{\prime\prime}$ components from
the requirement of a perturbative behaviour. We also investigate the
fixed point values of the top and $(B\!\!\!/)$ Yukawa couplings. We
discover that the former is somewhat reduced in strength from its
value in the $R_p$ conserving case while the latter are comparable
to the above-mentioned bounds.

RGE is our basic tool in constraining the couplings of all superfields
$\Phi^a$.  Here $a$ is a generic index.
Given a trilinear term $d_{abc} \Phi^a \Phi^b \Phi^c$ in the
superpotential and the evolution scale $\mu$, the RGE equation for
$d_{abc}$ is [9]
$$
\mu {\partial \over \partial\mu} d_{abc} = \gamma_a^{~i} d_{ibc} +
\gamma_b^{~j} d_{ajc} + \gamma_c^{~k} d_{abk}.
\eqno (3)
$$
In (3) $\gamma_a^{~i}$ is the anomalous dimension matrix
$Z^{-1/2~k}_{~~~~a}  \mu\partial/\partial\mu ~Z^{1/2~~i}_{~~~~k}$
where the renormalization constant $Z$ relates the renormalized
superfield $\Phi$ and the unrenormalized one $\Phi_0$ by
$$
\Phi_0^{~i} = Z^{1/2~i}_{~~~a} \Phi^a.
\eqno (4)
$$
We would apply (3) to the Yukawa couplings of interest.  The
Yukawa part of the Lagrangian density including
$(B\!\!\!/,R\!\!\!/_p)$ terms is [1]:
$$
{\cal L} = h_\tau [L_3 H_1 \bar E_3]_F + h_b [Q_3 H_1 \bar D_3]_F +
h_t [Q_3 H_2 \bar U_3]_F + \lambda^{\prime\prime}_{233} [\bar D_2 \bar
D_3 \bar U_3]_F + \lambda^{\prime\prime}_{133} [\bar D_1 \bar D_3 \bar
U_3]_F.
\eqno (5)
$$
As explained earlier, we include only the terms with the maximum
possible number of third generation indices in the $B\!\!\!/$ part of
${\cal L}$.  Of course, we also have the three regular Yukawa
terms coming from the third generation (characterized by the couplings
$h_t, h_b,h_\tau$) and ignore their lower generation counterparts.  In
(4) $H_1$ and $H_2$ are the Higgs superfields coupling to the up
and down quarks respectively.
$$
\vbox{
\begin{tabular}{|c|c|c|}
\hline
SUPERFIELDS & REPRESENTATIONS & ANOMALOUS DIMENSION \\
$\Phi$      & $SU(3)_L \times SU(2)_L \times U(1)_Y$ & $(4\pi)^2
\gamma^{~\Phi}_\Phi$ \\
\hline
$L_3$ & $(1,2,-1/2)$ & $h^2_\tau - \displaystyle{3 \over 2} g^2_2 -
\displaystyle {3 \over 10} g^2_Y$ \\
$\bar E_3$ & $(1,1,1)$ & $2h^2_\tau - \displaystyle{6 \over 5} g^2_Y$ \\
$\bar D_3$ & $(\bar 3,1,1/3)$ & $2h^2_b + 6\lambda^{\prime\prime
2}_{233} + 6\lambda^{\prime\prime 2}_{133} - \displaystyle{8 \over 3}
g^2_c - \displaystyle{2 \over 15} g^2_Y$ \\
$\bar D_2$ & $(\bar 3,1,1/3)$ & $6\lambda^{\prime\prime 2}_{233} -
\displaystyle {8 \over 3} g^2_c - \displaystyle {2 \over 15} g^2_Y$ \\
$\bar D_1$ & $(\bar 3,1,1/3)$ & $6\lambda^{\prime\prime 2}_{133} -
\displaystyle {8 \over 3}
g^2_c - \displaystyle {2 \over 15} g^2_Y$ \\
$\bar U_3$ & $(\bar 3,1,-2/3)$ & $2h^2_t + 6\lambda^{\prime\prime
2}_{133} + 6\lambda^{\prime\prime 2}_{233} - \displaystyle{8 \over 3}
g^2_c - \displaystyle {8 \over 15} g^2_Y$ \\
$Q_3$ & $(3,2,1/6)$ & $h^2_b + h^2_t - \displaystyle {8 \over 3} g^2_c
- \displaystyle {3 \over 2} g^2_2 - \displaystyle {1 \over 30}
g^2_Y$ \\
$H_1$ & $(1,2,-1/2)$ & $h^2_\tau + 3h^2_b - \displaystyle{3 \over 2}
g^2_2 - \displaystyle {3 \over 10} g^2_Y$ \\
$H_2$ & $(1,2,1/2)$ & $3h^2_t - \displaystyle{ 3 \over 2} g^2_2 -
\displaystyle {3 \over 10} g^2_Y$ \\
\hline
\end{tabular}
}
$$
\begin{center}
{\bf Table 1 : Properties of the relevent superfields.} \\
\end{center}

The $SU(3)_C \times SU(2)_L \times U(1)_Y$ representations and the
anomalous dimensions $\gamma_\Phi^{~\Phi}$ of the superfields
(diagonal elements of
the anomalous dimension matrix), occurring in (3), are given in Table
1 (N.B. $Q = I_3+Y$).  The $B\!\!\!/$ terms mix different generations
at the 1-loop level.  In particular,
they lead to nonzero values of the following
off-diagonal terms in the anomalous dimension matrices:
$$
\begin{array}{l}
\gamma_{\bar D_1}^{~\bar D_2} = \displaystyle {3 \over 8\pi^2}
\lambda^{\prime\prime}_{133} \lambda^{\star\prime\prime}_{233}, \\[2mm]
\gamma_{\bar D_2}^{~\bar D_1} = \displaystyle {3 \over 8\pi^2}
\lambda^{\star\prime\prime}_{133} \lambda^{\prime\prime}_{233}.
\end{array}
\eqno (5)
$$
All other off-diagonal elements vanish.

We can now write down the evolution equations of the Yukawa couplings.
(Soft supersymmetry breaking terms have dimensional strengths and they do not
contribute to the evolution of dimensionless couplings.)  The
equations are
$$
\begin{array}{l}
\mu \displaystyle {\partial \over \partial\mu} h_\tau = h_\tau
\left(\gamma_{L_3}^{~L_3} +
\gamma_{H_1}^{~H_1} + \gamma_{\bar E_3}^{~\bar E_3}\right), \\[2mm]
\mu \displaystyle{\partial \over \partial\mu} h_b = h_b
\left(\gamma^{~Q_3}_{Q_3} +
\gamma^{~H_1}_{H_1} + \gamma^{~\bar D_3}_{\bar D_3}\right), \\[2mm]
\mu \displaystyle{\partial \over \partial\mu} h_t = h_t
\left(\gamma^{~\bar U_3}_{\bar U_3} +
\gamma^{~H_2}_{H_2} + \gamma^{~\bar D_3}_{\bar D_3}\right), \\[2mm]
\mu \displaystyle{\partial \over \partial\mu} \lambda^{\prime\prime}_{133} =
\lambda^{\prime\prime}_{133} \left(\gamma^{~\bar U_3}_{\bar U_3} +
\gamma^{~\bar D_3}_{\bar D_3} + \gamma^{~\bar D_1}_{\bar D_1}\right) +
\lambda^{\prime\prime}_{233} \gamma^{~\bar D_2}_{\bar D_1}, \\[2mm]
\mu \displaystyle{\partial \over \partial\mu} \lambda^{\prime\prime}_{233} =
\lambda^{\prime\prime}_{233} \left(\gamma^{~\bar U_3}_{\bar U_3} +
\gamma^{~\bar D_3}_{\bar D_3} + \gamma^{~\bar D_2}_{\bar D_2}\right) +
\lambda^{\prime\prime}_{133} \gamma^{~\bar D_1}_{\bar D_2}.
\end{array}
\eqno (6)
$$
Next, we define $t = 1/(2\pi) \ell n(\mu/{\rm GeV})$ and, utilizing the
entries in Table 1, write (6) in terms of the $t$-evolution of
$\alpha_i$'s and $Y_i$'s, where $\alpha_i = g^2_i/(4\pi)$ and $Y_i =
h_i^2/(4\pi)$, $Y_{ijk} = \lambda^{\prime\prime 2}_{ijk}/(4\pi)$.  Thus
$$
\begin{array}{l}
\displaystyle{d\alpha_Y \over dt} = \left(\displaystyle{2} n_f +
\displaystyle{3 \over 5}\right)\alpha^2_Y, \\[2mm]
\displaystyle{d\alpha_2 \over dt} = \left(-\displaystyle{6} +
\displaystyle {2} n_f + \displaystyle{1}\right)\alpha^2_2, \\[2mm]
\displaystyle{d\alpha_3 \over dt} = \left(-9 + \displaystyle{2}
n_f\right) \alpha^2_3, \\[2mm]
\displaystyle{dY_\tau \over dt} = \left(4Y_\tau + 3Y_b + 3\alpha_2 -
\displaystyle {9 \over5} \alpha_Y\right)
Y_\tau, \\[2mm]
\displaystyle{dY_b \over dt} = \left(6Y_b + Y_t + Y_\tau + 6Y_{B\!\!\!/} -
\displaystyle{16\over3} \alpha_3 - 3\alpha_2 - \displaystyle{7\over15}
\alpha_Y\right) Y_b, \\[2mm]
\displaystyle{dY_t \over dt} = \left(6Y_t + Y_b + 6Y_{B\!\!\!/} -
\displaystyle{16 \over 3}
\alpha_3 - \displaystyle{13 \over 15} \alpha_Y\right)Y_t, \\[2mm]
\displaystyle{dY_{B\!\!\!/} \over dt} = \left(2Y_b + 2Y_t + 18Y_{B\!\!\!/}
- 8\alpha_3 - \displaystyle{4\over5} \alpha_Y\right) Y_{B\!\!\!/}.
\end{array}
\eqno (7)
$$
In (7), $n_f$ is the number of generations and we have defined
$Y_{B\!\!\!/}$ as the sum $Y_{133} + Y_{233}$ for convenience.

The low energy constraints on $Y_b,Y_t$ and $Y_\tau$ come from the
following relations:
$$
\begin{array}{l}
\sqrt{4\pi Y_t (m_t)} = m_t (m_t) \sqrt{1 +\tan^2\beta} (174~{\rm GeV}~
\tan\beta)^{-1}, \\[2mm]
\sqrt{4\pi Y_b (m_t)} = m_b (m_t) \sqrt{1 + \tan^2\beta}
(174 ~{\rm GeV}~\eta_b)^{-1}, \\[2mm]
\sqrt{4\pi Y_\tau (m_t)} = m_\tau (m_\tau) \sqrt{1 + \tan^2\beta}
(174 ~{\rm GeV}~\eta_\tau)^{-1}.
\end{array}
\eqno (8)
$$
In (8) $m_t (m_t)$, $m_b (m_b)$, $m_\tau (m_\tau)$ are the top, bottom
and tau masses at $\mu = m_t$, $\mu = m_b$, $\mu = m_\tau$
respectively.  We use $m_b (m_b) = 4.25 \pm .15$ GeV and $m_\tau
(m_\tau) = 1.777$ GeV as input values.  Moreover, with running
effects from loops taken into account [9], one may take $\eta_b \equiv
m_b (m_t)/m_b (m_b) = 1.54$ for $\alpha_s = .123 \pm .004$ and
$\eta_\tau \equiv m_\tau
(m_t)/m_\tau (m_\tau) \simeq 1$.  In our notation, $\tan\beta$ is the
ratio of vacuum expectation value of $H_2$ to that of $H_1$.  We know from
phenomenological analyses that [11] $\tan\beta$ lies between 1
and $m_t/m_b$.  Once the values of $Y_t$, $Y_b$ and $Y_\tau$
are fixed at the scale $m_t$, they can be evolved to the scale of SUSY
breaking $M_{SUSY}$ using non-SUSY RGE equations.  In our calculations
we have used $M_{SUSY}$ to be 1 TeV since this is preferred by the
data on gauge couplings for a unification scenario [2].  The top quark
mass range is taken to be 125-185 GeV.

We have not assumed any boundary conditions on the Yukawa couplings at
the unification scale.  This makes our analysis independent of models
of Yukawa coupling unification.  We only require that all Yukawa
coupling strengths remain perturbative $(Y_i < 1)$. (Later, we will
see the results in terms of the fixed points of these strengths).  In
our analysis this condition, when imposed, yields numerical upper
bounds on the magnitude of $\lambda_{B\!\!\!/}$, where
$\lambda^2_{B\!\!\!/}/(4\pi) = Y_{B\!\!\!/}$ (i.e.
$|\lambda_{B\!\!\!\!/}|
> |\lambda^{\prime\prime}_{133}|, |\lambda^{\prime\prime}_{233}|$) at
the scale $M_{SUSY}$.  The results have been collected in Table 2.

\bigskip

\begin{center}
$m_t ({\rm GeV}) \rightarrow$ \\
\end{center}
\vbox{$$
\begin{tabular}{|c|c|c|c|c|}
\hline
$\tan\beta$ & $125$ & $145$ & $165$ & $185$ \\
\hline
$1$ & & & & $Y_t$ \\
\hline
$5$ & $.63$ & $.60$ & $.56$ & diverges \\
\hline
$10$ & $.63$ & $.60$ & $.56$ & for any value \\
\hline
$20$ & $.62$ & $.58$ & $.56$ & of $Y_{133} + Y_{233}$ \\
\hline
$40$ & $.62$ & $.58$ & $.56$ & at the scale \\
\hline
$50$ & $.59$ & $.59$ & $.53$ & $M_{SUSY}$ \\
\hline
\end{tabular}
$$}
\begin{center}
{\bf Table 2 : Upper bounds on $\lambda_{B\!\!\!/}$ as functions of
$\tan\beta$ and $m_t$.} \\
\end{center}

We can highlight the following points regarding the above results.

\begin{enumerate}

\item {} Though we have taken $M_{SUSY} \simeq 1$ TeV, the value
preferred for gauge coupling unification [2], our bounds are
insensitive to the magnitude of the supersymmetry breaking scale.  The
choice $M_{SUSY} \simeq m_t$ yields more or less the same bounds with
some changes in the second decimal place.

\item {} An interesting point to note is that, towards higher values of
the top mass, the top coupling $Y_t$ itself diverges in the high $t$
region and the upper bound on the baryon number violating
couplings cannot be derived.

\item{} The constraint of perturbative unitarity at $M_{SUSY}$ would
directly imply that $Y_{B\!\!\!/} (M_{SUSY}) < 1$, i.e. $\lambda_{B\!\!\!/}
(M_{SUSY}) < 3.54$.  The dramatic effect of RGE and of the constraint
$Y_{B\!\!\!/} (\mu) < 1$ for $\mu < 2 \times 10^{16}$ GeV is to reduce
this upper bound from 3.54 to $0.5 - 0.6$.  Thus our achievement here,
after using RGE, is the reduction of the naive upper bound by a factor of
six or seven.

\item{} One can try to introduce a further restriction by requiring
[12] the unification of the three baryon number conserving Yukawa
couplings.  Here we note that the $B\!\!\!/$ Yukawa terms in the
superpotential have a different matter parity from the $B$ conserving
ones.  Therefore, even at the unification scale, there is no reason to
expect the equality of the $B$-conserving and $B\!\!\!/$ couplings.
However, the requirement of the unification of $Y_t,Y_b$ and $Y_\tau$
at that high scale mainly pushes the value of $\tan\beta$ towards the
higher region.  We see from Table 2 and Fig.1 that these bounds are
contained in our result.  In fact, the upper bounds do not vary much
with $\tan\beta$ when the latter is large.  However, $Y_b$ and
$Y_\tau$ start contributing substantially to the evolution of $Y_{B\!\!\!/}$
for higher values of $\tan\beta$. This reduces the upper
bound somewhat.

\item{} Ours is only a 1-loop RGE analysis.  Moreover, we have not
included the threshold effects here since they are typically of the
order of 2-loop terms [13] which have been ignored.  Since ours is the
initial investigation in this direction, we have tried to be
approximate and simple, rather than very accurate and complicated.

\end{enumerate}

There have been a number of studies [14] on the fixed point behaviour of
the top quark Yukawa coupling $Y_t$ in the MSSM.  It has been found
from the RGE that there is a fixed point value for $Y_t (m_t)$
starting from a large range of $Y_t$-values at the unification scale.
With $B\!\!\!/,R\!\!\!/_p$ couplings present however, two questions
arise automatically:

\begin{enumerate}

\item[{$\bullet$}] What happens to the fixed point of $Y_t$ now?

\item[{$\bullet$}] Is there a fixed point in $Y_{B\!\!\!/}$?

\end{enumerate}

\noindent We find that the evolutions of $Y_t$ and $Y_{B\!\!\!/}$ are
mutually dependent in a way that the fixed point in $Y_t$ is reduced
from the value it has in the MSSM.  First, we have calculated the
approximate fixed point of the Yukawa couplings by taking arbitrarily
large values of $Y_t$ and $Y_{B\!\!\!/}$ at the unification scale.
Such a scenario is possible in the SU(5) GUT where, at the unification
scale, one has the relation $h_\tau = h_b \neq h_t$.  It turns out that,
in the presence of $B$-violating Yukawa couplings, the fixed point of
$h_t$ reduces from 1.06 in the MSSM to 0.88.  We simultaneously get a
fixed point value of the $B\!\!\!/$ coupling $\lambda_{B\!\!\!/}$
around 0.59.  On the other hand, we can make all $Y$'s including
$Y_\tau$ and $Y_b$ arbitrarily large at the unification scale.  This
scenario may appear in the SO(10) GUT in the presence of a $16 \times
16 \times 10$ coupling at the GUT scale.  In this case the fixed point
value of $h_t$ reduces from 1.0 in the MSSM to nearly 0.86 in the presence of
$B$-violation while the fixed point value of $\lambda_{B\!\!\!/}$
is about 0.65.

Let us summarize our conclusions regarding constraints on
supersymmetric $B$-violating couplings.  Upper bounds have been derived on the
magnitudes of the
generation-hierarchically largest $B\!\!\!/$ Yukawa
couplings, possible in the MSSM extended to include $R_p$- and
$B$-violation, at the SUSY breaking scale.  These are not very
strong but are the first acceptable bounds on these hitherto
unresticted couplings.  Moreover, the present work treats
the evolution of $R\!\!\!/_p,B\!\!\!/$ coupling strengths
and their fixed point behavior in relation to that of the top Yukawa
coupling.   These bounds have emerged from the requirement of the
validity of perturbative behavior at all energies below the
unification scale.  Similar considerations can also be applied
to the other viable $R\!\!\!/_P$ model with the violation of lepton (but
not baryon) number.  However, in that case the quadratic self-driving
term would be smaller due to the absence of any color factor in the
1-loop graphs containing leptons.  So we expect less strong constraints
there.

B.B. would like to acknowledge the hospitality of the Tata Institute
of Fundamental Research and thank Prof. D.P. Roy for an initial
suggestion towards this work. P.R. would like to acknowledge the
hospitality of the
Physical Research Laboratory.  This paper was completed at the workshop
WHEPP3 (January 10-22, 1994) held at the Institute of Mathematical
Sciences, Madras, with joint sponsorship by the S.N. Bose National
Centre for Basic Sciences.

\newpage

\noindent {\bf References} \\

\begin{enumerate}

\item[{[1]}] L.J. Hall and M. Suzuki, Nucl. Phys. {\bf B231}, 419
(1984), S. Dawson, Nucl. Phys. {\bf B261}, 297 (1985), V. Barger,
G.F. Giudice, T. Han, Phys. Rev. {\bf D40}, 2987 (1989). L.J.
Hall, Mod. Phys. Lett. {\bf A5}, 467 (1990).

\item[{[2]}] U. Amaldi et. al., Phys. Lett. {\bf B260}, 447
(1991).

\item[{[3]}] H.P. Nilles, Phys. Rep. \underbar{C110}, 1 (1984).  P.
Nath, R. Arnowitt and A.H. Chamseddine, {\it Applied $N=1$ supergravity}
(World Scientific, Singapore, 1984).  H.E. Haber and G.L. Kane, Phys.
Rep. \underbar{C117}, 75 (1985).

\item[{[4]}] D. Brahm and L.J. Hall, Phys. Rev. {\bf D40}, 2449
(1989), L.E. Ib\'a\~nez and G.G. Ross, Nucl. Phys. {\bf B292}, 400
(1987).

\item[{[5]}] R. Barbieri and A. Masiero, Nucl. Phys. {\bf B267},
679 (1986).  A. Bouquet and P. Salati, Nucl. Phys. {\bf B284},
557 (1987).  B.A. Campbell, S. Davidson, J. Ellis and K. Olive, Phys.
Lett. {\bf B256}, 457 (1991).  J. Ellis and K. Olive, Phys. Lett.
{\bf B256}, 457 (1991).  E. Roulet and D. Tommasini, Phys. Lett.
{\bf B256}, 218 (1991).  W. Fischler, G. Giudice, R.G. Leigh and
S. Paban, Phys. Lett. {\bf B258}, 45 (1991).  A. Nelson and S.M.
Barr, Phys. Lett. {\bf B258}, 45 (1991).

\item[{[6]}] H. Dreiner and G.G. Ross, Nucl. Phys. {\bf B410}, 188 (1993).

\item[{[7]}] V. Barger et al, Ref. 1.
R. Godbole, P. Roy and X. Tata, Nucl.
Phys. {\bf B401}, 67 (1993).

\item[{[8]}] F. Zwirner, Phys. Lett. {\bf B132}, 103 (1983).

\item[{[9]}] P. West, ``{\it Introduction to supersymmetry and
supergravity}'' (World Scientific, Singapore, 1990).

\item[{[10]}] V. Barger, M.S. Berger and P. Ohmann, Phys. Rev.
{\bf D47}, 1093 (1993).

\item[{[11]}] Z. Kunszt and F. Zwirner, Nucl. Phys. {\bf B385}, 3 (1992).

\item[{[12]}] B. Ananthanarayan, G. Lazarides, Q. Shafi, Phys. Rev.
{\bf D44}, 1613 (1991).

\item[{[13]}] J. Ellis, S. Kelley, D.V. Nanopoulos, Nucl. Phys.
{\bf B373}, 55 (1992); P. Langacker and N. Polonsky, Phys. Rev.
{\bf D47}, 4028 (1993).

\item[{[14]}] B. Pendleton and G. G. Ross, Phys Lett {\bf B98}
291 (1981); C. T. Hill, Phys. Rev {\bf D24}, 691, (1981). For a
brief summary of RGE results for SUSY GUTS see V. Barger, M.
S. Berger, P. Ohmann and R. N. J. Phillips, University of
Wisconsin Madison report, MAD/PH/803 and references therein.

\end{enumerate}

\newpage

\noindent \large {\bf Figure caption} \\

\begin{enumerate}

\item[{\rm Fig.~1}] Three curves showing the variation on the upper
bound on $\lambda_{B\!\!\!/}$ with respect to the free parameter
$\tan\beta$ for three different values (in GeV) of the top mass.

\end{enumerate}

\end{document}